# The Terahertz Intensity Mapper (TIM): a Next-Generation Experiment for Galaxy Evolution Studies

Joaquin Vieira, James Aguirre, C. Matt Bradford, Jeffrey Filippini, Christopher Groppi, Dan Marrone, Matthieu Bethermin, Tzu-Ching Chang, Mark Devlin, Oliver Dore, Jianyang Frank Fu, Steven Hailey Dunsheath, Gilbert Holder, Garrett Keating, Ryan Keenan, Ely Kovetz, Guilaine Lagache, Philip Mauskopf, Desika Narayanan, Gergo Popping, Erik Shirokoff, Rachel Somerville, Isaac Trumper, Bade Uzgil, and Jonas Zmuidzinas

*Abstract*— Understanding the formation and evolution of galaxies over cosmic time is one of the foremost goals of astrophysics and cosmology today. The cosmic star formation rate has undergone a dramatic evolution over the course of the last 14 billion years, and dust obscured star forming galaxies (DSFGs) are a crucial component of this evolution. A variety of important, bright, and unextincted diagnostic lines are present in the far-infrared (FIR) which can provide crucial insight into the physical conditions of galaxy evolution, including the instantaneous star formation rate, the effect of AGN feedback on star formation, the mass function of the stars, metallicities, and the spectrum of their ionizing radiation. FIR spectroscopy is technically difficult but scientifically crucial. The FIR waveband is impossible to observe from the ground, and spans a crucial gap in the spectroscopic coverage between the Atacama Large Millimeter/submillimeter Array (ALMA) in the sub/mm, and the James Webb Space Telescope (JWST) in the mid-IR. Stratospheric balloons offer a platform which can outperform current instrument sensitivities and are the only way to provide large-area, wide bandwidth spatial/spectral mapping at FIR wavelengths.

NASA recently selected TIM, the Terahertz Intensity Mapper, with the goal of demonstrating the key technical milestones necessary for FIR spectroscopy. TIM will provide a technological steppingstone to the future space-borne instrumentation such as the Origins Space Telescope (OST, formerly the Far-IR Surveyor) or a Probe mission. TIM will address the two key technical issues necessary to achieve this:

1. Low-emissivity, high-throughput telescope and spectrometer optics for the FIR;
2. Background-limited detectors in large format arrays, scalable to >10,000 pixels.

We will do this by constructing a integral-field spectrometer from 240 - 420 microns with 3600 kinetic-inductance detectors (KIDs) coupled to a 2-meter low-emissivity carbon fiber telescope.

In addition to the development and demonstration of crucial technologies for the FIR, TIM will perform groundbreaking science. We will survey two 0.1 square degree fields centered on GOODS-S and the South Pole Telescope Deep Field, both of which have rich ancillary data. Scientifically, we will:

1. Obtain spectroscopic line detections of ~100 galaxies in the atomic fine structure lines [CII] (158 microns) (at $0.5<z<1.5$), [NII]

Manuscript submitted June 25, 2019.
J. Vieira is with the Department of Astronomy, University of Illinois Urbana-Champagne, Urbana IL 61801 USA (email: jvieira@illinois.edu).
J. Aguirre is with the Department of Physics and Astronomy, University of Pennsylvania, Philadelphia, PA 19104 USA (email: jaguirre@sas.upenn.edu).
C.M. Bradford is with the NASA Jet Propulsion Laboratory, Pasadena CA 91109 USA (email: bradford@submm.caltech.edu)
J. Filippini is with the Department of Astronomy, University of Illinois Urbana-Champagne, Urbana IL 61801 USA (email: jpf@illinois.edu).
C.E. Groppi is with the School of Earth and Space Exploration, Arizona State University, Tempe, AZ 85287 USA (e-mail: cgroppi@asu.edu).
D. Marrone is with Steward Observatory, University of Arizona, Tucson, AZ 85721 USA (email: dmarrone@email.arizona.edu).
M. Bethermin is with the Laboratoire d'astrophysique de Marseille, 13388 Marseille FRANCE (email: matthieu.bethermin@lam.fr).
T-C. Chang is with the NASA Jet Propulsion Laboratory, Pasadena CA 91109 USA (email: tzu-ching.chang@jpl.nasa.gov)
M. Devlin is with the Department of Physics and Astronomy, University of Pennsylvania, Philadelphia, PA 19104 USA (email: devlin@physics.upenn.edu).
O. Dore is with the NASA Jet Propulsion Laboratory, Pasadena CA 91109 USA (email: olivier.p.dore@jpl.nasa.gov).
J.F. Fu is with the Department of Astronomy, University of Illinois Urbana-Champagne, Urbana IL 61801 USA (email:fu32@illinois.edu).
S. Hailey Dunsheath is with the Department of Astronomy, California Institute of Technology, Pasadena, CA 91125 USA (email: haileyds@caltech.edu).

G. Holder is with Department of Astronomy, University of Illinois Urbana-Champagne, Urbana IL 61801 USA (email: gholder@illinois.edu).
G. Keating is with the Harvard Smithsonian Center for Astrophysics, Cambridge, MA 02138 USA (email: garrett.keating@cfa.harvard.edu).
R. Keenan is with the Steward Observatory, University of Arizona, Tucson, AZ 85721 USA (email: rpkeenan@email.arizona.edu).
E. Kovetz is with the Department of Physics and Astronomy, Johns Hopkins University, Baltimore, MD 21218 USA (email: elykovetz@jhu.edu).
G. Lagache is with the Laboratoire d'astrophysique de Marseille, 13388 Marseille FRANCE (email: guilaine.lagache@lam.fr).
P. Mauskopf is with the School of Earth and Space Exploration, Arizona State University, Tempe, AZ 85287 USA (e-mail: Philip.Mauskopf@asu.edu).
D. Narayanan is with the Department of Astronomy, University of Florida, Gainsville, Gainsville, FL 32611 USA (email: desika.narayanan@ufl.edu).
G. Popping is with the Max Planck Institute for Astronomy, 69117 Heidelberg, Germany (email: popping@mpia.de).
E. Shirokoff is with the Department of Astronomy and Astrophysics, University of Chicago, Chicago, IL 60637 USA (email: shiro@chicago.edu).
R. Somerville is with the Department of Physics and Astronomy, Rutgers University, Piscataway, NJ 08854 USA (email: somerville@physics.rutgers.edu).
I. Trumper is with Steward Observatory, University of Arizona, Tucson, AZ 85721 USA (email: itrumper@optics.arizona.edu).
B. Uzgil is with the Max Planck Institute for Astronomy, 69117 Heidelberg, Germany (email: uzgil@mpia.de).
J. Zmuidzinas is with the Department of Astronomy, California Institute of Technology, Pasadena, CA 91125 USA (email: jonas@caltech.edu).

(205 microns) (at 0.2<z<1 ), [OI] (63 microns) (at 2.8<z<5.7 ) and [OIII] (88 microns) (at 1.7<z<3.8);

2. Establish the mean star formation rate (proportional to [CII] luminosity), metallicities (proportional to the [CII]/[NII] ratio), and AGN content (proportional to the [OIII] luminosity) of galaxies using a stacking analysis of known sources in the field;

3. Produce deep maps of the 3D structure of the Universe by redshift tomography ("intensity mapping") with [CI], and [CII] × [NII] cross-spectra, to constrain the cosmic star formation history at cosmic noon and lay the important groundwork for extending this technique to even higher redshifts to eventually explore the epoch of reionization.

In this paper, we will summarize plans for the TIM experiment's development, test and deployment for a planned flight from Antarctica in Austral summer of 2022-2023.

*Index Terms*—Astronomy, suborbital, balloon, intensity mapping, kinetic inductance detector.

I. INTRODUCTION

Explaining the history of cosmic star formation through the evolution of galaxies is one of the most important challenges in modern astrophysics. A wealth of data has been assembled in the last decade showing clearly that the total star formation rate density has fallen dramatically since its peak 7−10 Gyr ago (z~1-3; [1]). The nature of the galaxies responsible for the bulk of star formation has also changed of cosmic time, with star formation previously dominated by luminous, dust-obscured, star-forming galaxies (DSFGs) that are almost absent in the local universe [2].

The path to understanding galaxy evolution will necessarily run through observations in the far infrared (FIR). Half of the total energy output from the cosmic star formation has been absorbed by interstellar dust and re-emitted in the [3],[4]. Moreover, there are a variety of un-extincted FIR diagnostic lines that can reveal the physics of galaxy evolution by tracing the star formation rate (SFR), black hole accretion rate, mass function of stars, spectrum of ionizing radiation, and metallicity of the interstellar medium (ISM).

Spectroscopy in the FIR is technically difficult but essential to the study of galaxy evolution. We are building the Terahertz Intensity Mapper (TIM), with the goal of demonstrating balloon-borne FIR spectroscopy limited by the photon noise from the atmosphere. TIM will be a vital technological, data analysis, and scientific steppingstone to future orbital missions, and will also advance our understanding of galaxy evolution through observations that cannot be replicated with current FIR instruments. TIM combines a long-slit spectrometer operating from 240−420 μm with a 2 m low-emissivity carbon-fiber telescope to provide a substantial increase in sensitivity over existing instruments. We will survey one 0.1 deg$^2$ field centered on GOODS-S and one wider field (~1 deg$^2$) within the South Pole Telescope (SPT) Deep Field, both of which have rich multi-wavelength ancillary data.

The science goals of TIM are:

1. Produce deep tomographic maps of the 3D structure of the Universe to measure the power spectrum of [CII] and [CII]×[NII]. This will be a pioneering demonstration of the technique of "intensity mapping," which provides a new method to constrain the cosmic star formation history and measure its relation to the underlying dark matter distribution;
2. Perform a blind spectroscopic survey for [CII] line emitters within an enormous cosmic volume, $10^7$ Mpc$^3$, at 0.52<z<1.67. We expect to detect ~100 galaxies, which will be a powerful observational constraint on models of galaxy evolution.
3. Capture the star formation contribution of galaxies too faint to be detected individually, by measuring the [CII] luminosity function across the peak of cosmic star formation;
4. Use stellar mass-selected galaxies with spectroscopic redshifts from the GOODS-S field to stack on [CII] and [NII], and develop the theory to relate this to the total star formation rate ([CII]), star formation mode ([CII]/$L_{FIR}$), metallicity ([NII]/[CII]), and specific star formation rate ([CII]/$M_{star}$);
5. Cross-correlate the [CII] data cube (which provides redshift information) with Herschel/SPIRE maps (SFR) to calibrate the [CII]/SFR relation, and Spitzer/IRAC maps (stellar mass) to measure the specific star formation rate versus redshift.

TIM is a wholly unprecedented experiment to study the cosmic star formation history. It will map a volume spanning 4.5 billion years of cosmic history (0.52<z<1.67), on scales from 1−50 Mpc (30" to ~1°) with complete spectroscopic information. In the coming decades this will be a powerful cosmological tool for charting the 3D structure of the universe. There is significant discovery potential with TIM, since it will be probing an under-explored wavelength range with unprecedented sensitivity using a new astrophysical technique. TIM will be the first generation of experiments using intensity mapping in the FIR regime and fills in a crucial wavelength regime only accessible from either space or a balloon platform. TIM fills a unique and vital scientific niche not filled by Herschel, SOFIA, ALMA, JWST, or even SPICA as currently conceived.

II. INSTRUMENT

*A. Design Considerations and Sensitivity*

To achieve the scientific and technical goals of TIM, we must be able to demonstrate atmosphere limited performance of the telescope and detectors. Photon loading this low is only possible from a (sub)orbital platform and motivates the necessity of a balloon program. To estimate the atmosphere background, the (proprietary) ATM model of Juan Pardo was used. We have assumed a flight at mid-latitudes, an altitude of 37 km, and observations at 45" elevation. The ATM model calculates the opacity due to all relevant atmospheric species. We calculate the noise equivalent power (NEP) due to photon noise from a greybody in the usual manner assuming an instrument transmission of 25% and two photon modes (both polarizations, horn coupled) and require that the intrinsic noise in the detectors be sub-dominant to this photon noise. The line sensitivity on the sky then incorporates the point-source efficiency, ~64% with the horn architecture we choose.

We have considered the performance of various potential telescope architectures for TIM, incorporating the atmospheric transmission and loading, a range of telescope sizes, and the possibility of actively cooling the telescope to minimize its thermal emission. While a cooled aperture performs better, it is very costly for a given aperture size, and the performance improvement is modest because even at balloon altitudes there is ~1% emission from the 250 K atmosphere. In light of these calculations, we believe we can demonstrate near atmosphere-limited performance using the on-axis telescope described below with carefully controlled primary illumination, cold stops at pupils in the system, and baffling to prevent spillover to warm surfaces.

In principle, the required spectral resolution and large area mapping could be achieved with either a Fabry-Perot (FP) or FTS spectrometer design. However, both of these incur sensitivity penalties. The FP does not cover the entire frequency range instantaneously and must be scanned, and the FTS places the full optical bandwidth of the entire band on each detector, increasing the noise. The best approach is a reflective, blazed diffraction grating, which orders large instantaneous bandwidth and good sensitivity. We target a spectral resolving power $R = 250$ (830 kms$^{-1}$), a value that is well-suited to intensity mapping because it enables instantaneous coverage of a wide band (32%) with a single spectrometer with a modest number of resolution elements (<100). While $R = 250$ under-resolves galaxies' intrinsic line widths, the sensitivity degradation is only modest ($\propto \sqrt{\delta\nu}$) for a photon-noise limited instrument, and the wider bandwidth recovers this factor through the speed increase for blind survey experiments, as demonstrated with Z-Spec (e.g. [5]). The instrument parameters and resulting sensitivity for our design are given in Table 1.

*B. Telescope*

The TIM telescope design must be lightweight and compact, with low overall emissivity (i.e., any coupling to warm, non-astrophysical radiation loads) and high efficiency coupling to the spectrometer. In principle, an off-axis telescope can achieve a lower emissivity than an on-axis one, due to the lack of scattering from the secondary mirror and its support legs. However, off-axis mirrors are more costly and present fabrication challenges. We have designed a segmented, 2.0-meter diameter, carbon fiber primary mirror, as well as a fully carbon fiber secondary and support structure, with gold metallization of the reflecting surfaces to minimize the emissivity.

There are three key specifications that lead to our current design. First, we are driven to the largest aperture we can accommodate by sensitivity and resolution requirements, particularly for source stacking and single-object detections in our survey fields. The mirror diameter defines the resolution and therefore the maximum wavenumber (k) accessible to the experiment. A 2.0-meter mirror is the minimum size at which we can study both the clustered and Poisson portions of the power spectrum and should also retain appreciable signal to noise for stacking. Second, our telescope must be both cost and weight efficient. The BLAST aluminum 1.8-meter primary mirror had a surface mass density of 45 kg/m$^2$; this would be 110 kg for the mirror alone for a 2.0-meter aperture. Our carbon fiber reinforced polymer (CFRP) mirror will be 7-10 kg/m$^2$, making the telescope a marginal contributor to the mass and moment of inertia of the payload. Fabrication of a 2.0-meter mirror poses significant challenges, however. FIR-quality mirrors are usually machined from aluminum on diamond lathes, but there are very few machines that can produce a mirror this large. Even CFRP, which is assembled rather than machined, becomes costlier and riskier at this large diameter. For this reason, we have elected to divide the mirror into 6 segments, reducing the fabrication scale to 1.0 m where we have access to key components (e.g., oven, metallization chamber, machines for CF pressing) of appropriate size. The third constraint, low emissivity, drives the requirement of a gold reflective surface. The emissivity of gold is half of that of aluminum, which will provide a useful sensitivity boost. The total system emissivity will also be lowered by using three feedlegs to support the secondary (as in the current BLAST-TNG design, but unlike the four in BLAST), and by covering the feedlegs with reflective baffles to direct light incident on the legs to the cold sky. The final expected emissivity is 2-2.5%, compared to 4% in measured by BLAST. We assume 2.5% emissivity for sensitivity calculations, as given in Table 1.

The telescope concept has been prepared in collaboration with Composite Mirror Applications (CMA), a Tucson company that specializes in CFRP reflectors for optical wavelengths. CMA has demonstrated highly repeatable replication of spherical and paraboloidal optical mirrors, with surface figure accuracies of ~100 nm (50-100× better than needed) for mirrors as large as 0.8 m (Figure 1). For TIM wavelengths, accuracies of a few microns will be more than adequate, and there are many published examples of meter-scale CFRP mirrors by CMA and other vendors that exceed this requirement (e.g., [6], [7], [8]). The CFRP mirror segments will be made by replicating against a glass mandrel, whose precision should determine the accuracy of the mirror. The mandrel for this project will be cast, ground and polished to the appropriate figure using the tools developed for making large optical mirrors. The mirror segments will be built on top of the mandrel by laying unidirectional CFRP sheets in many orientations and curing under pressure. Each individual mirror segment will be subjected to precision metrology at the University of Arizona, using laser trackers and the Software Configurable Optical Test

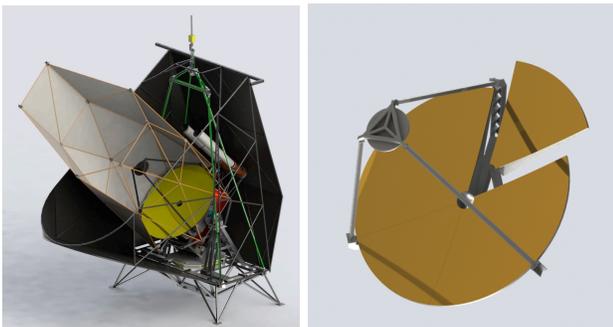

Fig. 1: Left: TIM telescope and gondola, showing the cryostat and star cameras. We will re-use the design for the gondola, cryostat, readout electronics, and star camera design from BLAST-TNG, which have all been constructed and are being prepared for a flight in late 2018. Right: The design for the segmented CFRP mirror from UA and CMA.

System [9] to confirm their figure. The six segments, after

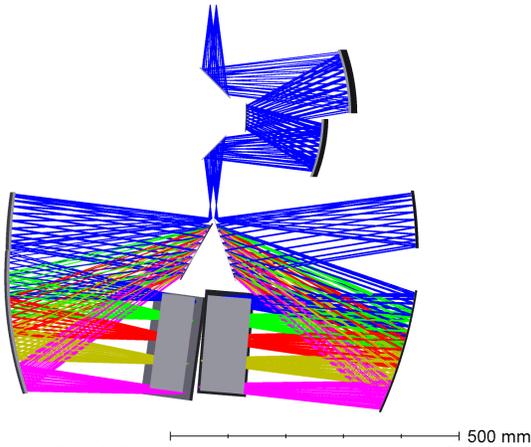

Fig. 2: TIM optical design. Two spectrometer modules are fed with a common Offner-style relay that provides a cold pupil mirror. Pickoff mirrors placed directly behind the slits redirect the light the individual modules. The full package is sized for the BLAST-TNG cryostat.

gilding, will be aligned and pinned at Arizona using similar validation tools.

The entire telescope, including the structural support for the secondary mirror feedlegs, which will anchor to the rear hub at the center of the primary mirror, will be made from CFRP. The operating temperature of this structure is not very low (250 K), and thermal contraction for CFRP is 20× smaller than aluminum, but matching materials in this way should lead to conformal contraction of the telescope (to first order). Residual correction can be made by focusing the telescope upon reaching flight altitude, where BLAST has observed just ±1.5 K diurnal variations. We will use the BLAST focusing system which has three precision actuators behind the secondary to provide 3-micron positioning. The entire system worked flawlessly during the BLAST 2006, 2010, and most recently 2012 [10], [11]. The secondary focus is located in the same location as for BLAST-TNG to allow re-use of the gondola and inner frame design, also shown in Figure 5. The maximum field of view of the telescope (defined as when the beam at 300 μm drops to a Strehl ratio of 0.95) is 0.5" in diameter. This is well-matched to TIM, and the design could be reused for other imaging submillimeter balloon missions.

### C. Spectrometer Architecture

To efficiently perform spectroscopy over the full 240 − 420 μm band, we will deploy two independent spectrometer modules: a short wavelength (SW) module covering 240−317 μm, and a long wavelength (LW) module covering 317 − 420 μm. The two modules follow the same basic design; each consists of a plane diffraction grating mounted between concave collimating and camera mirrors in a Czerny- Turner configuration. The optics are designed to accept an f/4 cone at the entrance slit and produce a telecentric f/4 image at the output. The grating is operated in first order and is sized to provide a resolving power of R ≈ 250. The diffraction gratings will be cut from M1 mold plate and machined with diamond tooling, similar to ZEUS-2 [12], [13]. The linear dimensions of the SW module are a factor of 1.3 times smaller than those of the LW module.

The full optical layout is shown in Figure 2. A set of cold relay optics (M3 − M6) reimages the f/5 telescope focus to an f/4 image at the entrance slits of the spectrometer modules. An image of the secondary mirror is formed on M4, and this optic serves as a Lyot stop to block stray radiation. The central circular area of M4 conjugate to the hole in the primary mirror will also be painted black to further reduce the optical loading. The LW and SW slits are 350 and 260 in length, respectively, are aligned in azimuth, and separated by 2.60.

Pickoff mirrors placed directly behind the entrance slits direct the radiation away from the common optical axis, allowing the spectrometer modules to be well separated. Another set of pickoffs (SW3 and LW3) placed after the camera mirrors position the focal planes adjacently, allowing the two detector modules to be mounted in close proximity on a sub-Kelvin stage. The full set of optics occupies a 770 mm × 610 mm × 350 mm volume and is sized to approximately match the instrument volume available in the BLAST-TNG cryostat.

The spectrometer modules are designed for modularity to allow separate testing and optimization in a test cryostat smaller than the full TIM cryostat.

### D. Array format, Horn Coupling

Each focal plane will be sampled by an array of 1800 hexagonally close-packed, straight-walled conical feed horns (see Figure 3). We utilize a 25 (spatial) × 72 (spectral) element array of 1.5Fλ horns (though the optical design above accommodates ~42 spatial beams). The Strehl ratio over much of this focal plane is (0.90, with some degradation at the array edges. The optics provide an instantaneous spectral coverage of $\lambda_{max}/\lambda_{min} \approx 32\%$ at R ~250.

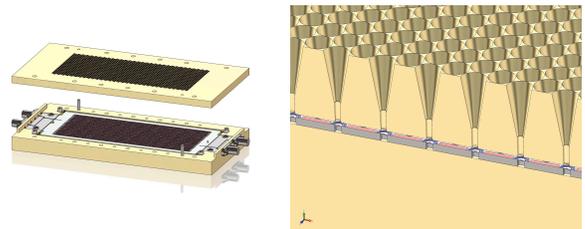

Fig. 3: TIM horn coupled focal-plane array architecture. Conical multi-flare-angle horns are drilled into the metal substrate with a custom tool. Alignment between the metal and the silicon die is via a pin and a pin and-slot, accommodating the differential CTE.

While optimized for intensity mapping, the spectrometer is sensitive to individual galaxies. The width of the entrance slit impacts the monochromatic point source sensitivity by controlling the amount of background power seen by a detector, as well as the fraction of the point source radiation entering the spectrometer. For a 1.5Fλ horn this sensitivity is maximized for a slit width of ≈Fλ, which yields a point source coupling efficiency of ≈0.64. Most of the rest of the horn power terminates on the entrance slit or the Lyot stop, with only a small loss in point source coupling due to a finite beam taper at the edge of the pupil. The spectrometer optics produce a small amount of anamorphic magnification, but this has negligible effect on the point source coupling.

The spectrometer modules are housed in separate 1K enclosures, with blackened surfaces and optical baffling control stray light and minimize loading on the detectors. A set of IR blocking filters at 77K, 40K, and 4K is used to reduce the loading on the cold stages, and a final set of capacitive low-pass metal mesh filters mounted to the horn arrays are used to define the bandpass of each module. We estimate the optical efficiency of the cold instrument, including the transmission and illumination efficiencies of all filters and mirrors, as well as the finite point source coupling provided by the horns, at 25%.

### E. Kinetic-Inductance Detector (KID) Arrays

KIDs have emerged in the last decade as a straightforward approach to very large detector arrays for astrophysics. These devices use thin-film, high-Q micro-resonators that absorb incident radiation and respond by changing resonance frequency and linewidth. Due to the high resonance quality factors $Q \sim 105 = f/\delta f$ can be obtained (corresponding to narrow line widths), large numbers of KIDs may be read out on a single RF/microwave circuit, and the only cryogenic electronics necessary is a single cold (4–20 K) RF/microwave amplifier per readout circuit. Each circuit is simply a single RF line down to the focal plane and another line returning via the amplifier, and it carries the signals of ∼$10^3$ detectors.

KID technology is now rivaling the performance levels of the SQUID-multiplexed bolometer systems in ground-based instruments. One example is the dual-band 150/240 GHz, 2896-pixel NIKA-2 camera fielded at the IRAM 30-m telescope by European groups at SRON Utrecht (Baselmans), Institut NEEL (Benoit), and Cardiff (Doyle), which has demonstrated sensitivities approaching the photon background limit. KIDs are being delivered now for flight with BLAST-[14], [15] and

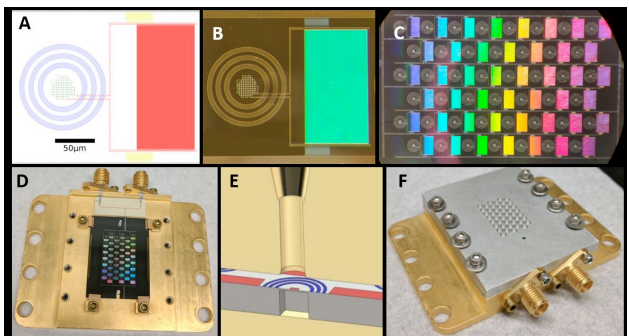

Fig. 4: (A) Diagram of the mask layout for a single TIM pixel. The meandered inductor (green) is surrounded by an optical choke structure (blue). An interdigitated capacitor (red) sets the resonance frequency of the pixel, and two coupling capacitors (yellow) couple microwave signal onto microstrip feedlines. (B) A microscope image of a single pixel as fabricated. All pixel elements of the prototype array are patterned out of 40 nm Al film. (C) A microscope image of the 45-pixel prototype array, as fabricated. (D) The fabricated array in its enclosure. The back side of the die is bare silicon and lies flat on the gold-plated package surface. The full size of the die is 30mm×22mm. (E) A CAD model of the detector package. The optical power is coupled into a feedhorn, and travels through a circular waveguide that is terminated by the inductor of the LEKID. A backshort is formed by deep trench etching from the backside to a 27-micron buried oxide layer, then metallizing. (F) The prototype feedhorn block installed above the 45-pixel array.

are under development for the TolTEC mm-wave camera to be deployed to the Large Millimeter Telescope (LMT). At Caltech / JPL, we have fielded the 350-micron camera MAKO [16],

[17] at the CSO before its shutdown. More recently, we have demonstrated high yield and detector sensitivity below $10^{-18}$ W $Hz^{-1/2}$ in our mm-wave on-chip spectrometer SuperSpec ([18], [19], [20], [21]). A SuperSpec demonstration instrument is being prepared now for deployment to the LMT in the winter of 2018–2019.

Achieving photon-noise-limited performance on the TIM balloon-borne spectrometer requires more sensitive KIDs than have yet been fielded scientifically. We have designed the TIM pixel, prototyped small arrays in the JPL microdevices laboratory, and demonstrated detector sensitivities which outperform the TIM photon noise specification by a factor of ∼2.5. The approach is shown in Figure 4, and top-level parameters are provided in Table 1. We employ the same basic single-layer lumped-element architecture as shown in Figure ??. We use a similar interdigitated capacitor design; this sets the two-level-system (TLS) frequency noise, ensuring that it is sub-dominant to the other noise sources. We use the same 100–250 MHz resonator frequencies, enabling us to use digitize directly in baseband without mixing. To meet the lower NEP required for TIM (as well as the other low-background applications), we need only reduce the volume of the inductor to increase the power-to-fractional-frequency responsivity, $R = (df/f)/dP_{inc}$, which scales as 1/V. The low volume, together with the requirement to create an impedance-matched absorber from the meandered inductor drives us to aluminum rather than titanium nitride.

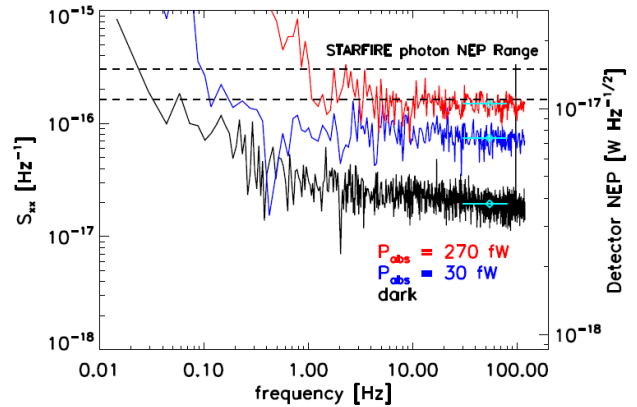

Fig. 5: Noise equivalent power (NEP) of a typical starfire KID on the prototype array shown in Figure 4. The left axis is fractional fractional frequency noise variance $S_{xx}$, the right is the NEP = $\sqrt{S_{xx}}/R$, where R is the power to fractional frequency responsivity ($1.2 \times 10^9 W^{-1}$ for these devices). The detector noise is dominated by photons for loading at and below that expected on float with TIM. The dark noise indicates a detector 1/f which produces knee at <0.1 Hz when compared with the expected photon noise.

The inductor volume is too small to permit a full MAKO-style 2-D meandered absorber, so each TIM detector will use a circular waveguide fed by a feedhorn (Figure 7), an approach broadly similar to that for super-BLAST-Pol [22], but with design enhancements to support our low-NEP device. The electromagnetic structure is designed to couple to both $TE_{11}$ modes in the circular guide, and the design includes a flare at the bottom of the waveguide and a lithographically patterned choke structure (3 concentric annular rings on the wafer surface) to help eliminate conversion into substrate modes. The backshort is integral to the device – it is created by simply etching from the backside to a buried oxide layer at the appropriate depth (27 μm for the long-wave device), then depositing aluminum.

At the heart of the device is the inductor / absorber—a single meander of 0.4-micron-wide aluminum patterned into the 40-nm-thick film with a total volume of 76 μm$^3$. The single meander couples as a 'mesh' to both polarizations by allowing the various segments of meander line to come close enough to one another at the corners to create capacitive shorts at the optical frequencies (715–1250 GHz). This requires a 0.5 μm gap and a 1.5 μm overlap length for each of the intersections. Our measurements, as well as those with the MAKO devices, indicate that this indeed couples well to both polarizations, and does not impact yield or readout frequency.

We have fabricated 45-pixel TIM prototype sub-arrays and matching feedhorn blocks shown in Figure 4. To demonstrate performance, we use our cryogenic test facility in which a cryogenic blackbody illuminates the array through with a 350-micron bandpass filter. The results are shown in Figure 5. We find that the devices meet the TIM target already, showing photon noise 2.5× better. Our characterization is based on the run of noise and responsivity with stage temperature and optical loading, as well as dark noise measurements. The observed optical efficiency of the prototype system is 20%, exactly as modeled in HFSS for this backshort-less design. The same HFSS models indicate 80% once the backshort is incorporated. Our characterization also shows that the system is well-explained by the superconductivity theory and the simple model for the quasiparticle recombination rates in the aluminum (see [23], [24]) The low-loading (dark) noise floor is due to generation-recombination (GR) of residual quasiparticles at the 220 mK operating temperature. We refer the reviewer to [25] for more details. The device yield in the current prototype is 89% (40/45), which exceeds our 80% requirement, but we expect to improve yield prior to TIM array delivery as we fine tune the fabrication recipe.

*F. Readouts*

In order to achieve frequency domain multiplexing of KID arrays, two tasks must be accomplished: 1) A waveform consisting of a sum of frequency tones (each at an individual pixel frequency) must be generated and transmitted to the array and 2) after interacting with the pixels, the complex transmission of the individual tones must be extracted from the waveform. The first task is easily accomplished using a cyclic memory buffer and a DAC to continuously play back a pre-calculated periodic waveform. The second task can be accomplished using advanced digital-signal processing hardware. A fast, high dynamic-range ADC is followed by a Field Programmable Gate Array (FPGA) to detect the phase and amplitude modulation of the signal tones by changes in the KID inductance.

TIM will use the system developed for BLAST-TNG. This system leverages the Reconfigurable Open Architecture Computing Hardware 2 (ROACH2) platform developed by the Berkeley CASPER group which features a Xilnix Virtex-6 FPGA. An additional daughter card provides two 1 GSPS DACs and two 500 MSPS ADCs. DRAM on the ROACH2 enables waveform playback by both DACs simultaneously. The TIM system is simpler in that we do not require mixing to reach the KID resonant frequency. The readout uses custom FPGA firmware developed specifically for KID readout. The firmware implements a polyphase filter bank (PFB) followed by a fast Fourier transform (FFT) to achieve an initial stage of coarse frequency separation into 1024 channels. This is followed by multiplication of the FFT channels by pre-specified sinusoids and then low-pass filtering. This readout is working in the laboratory and is being deployed with BLAST-TNG for its flight in late 2016 or early 2017.

The total noise from the cold amplifier and readout electronics is a comfortable factor of 10 lower (in $S_{xx}$ units) than the noise from the devices themselves. A single readout can perform simultaneous complex transmission measurements of 1024 tones at a rate of 488 Hz, plenty of margin on the 900 detectors per readout chain we intend to use with TIM. We require 4 ROACH2 readout chains reading out 1024 detectors each, 2 per spectrometer. Additionally, a readout computer is used to perform post-processing of the data and storage of the time stream. We budget two computers to service the ROACH2 systems, and we will use computers optimized for lower power consumption.

*G. Cryogenics*

The TIM cryostat and cryogenic design will copy BLAST-TNG, which builds from the flight heritage of BLAST and BLASTpol. The design is intended to hold for 28 days, based on the prototype, long enough for an LDB flight, with an exceptionally large cryogenic volume. Unlike previous designs, the BLAST-TNG cryostat uses only LHe in a single 250 L tank which forces the boiled-off He gas through two heat exchangers that cool two vapor-cooled shields, which operate at 66 K and 190 K. A separate pumped helium pot maintains a 1 K stage with 20 mW of cooling power, which contains the entire spectrometer. The detectors are cooled using pumped $^3$He/$^4$He sorption fridges which provide a 300 mK sink during flight with 30 μW of cooling power for 3 days. It is backed and cycled by a $^4$He stage. It can be recycled within 2 hrs.

*H. Gondola and Pointing System*

The TIM gondola and pointing system is designed around the successful BLAST heritage. The gondola is shown in Figure 5. It consists of a precision-pointed inner frame (composed of the primary, secondary, near-field baffle, and cryostat) supported by an external gondola. The outer frame is pointed in azimuth by a flywheel and an active pivot. The inner frame has an elevation mount with direct drive servo motors driving it relative to the outer frame. Balance of the inner frame is maintained by pumping liquid from the bottom of the frame to the top to compensate for cryogen boil-off.

The attitude determination system uses an array of pointing sensors including two sophisticated star cameras, two sets of fast, low-drift gyroscopes, a quad-GPS system, a digital Sun sensor, encoders, tilt sensors, and a magnetometer. The software is written to take full advantage of the abilities of each sensor in a hierarchical scheme where the fast, high-drift sensors (gyros) are continuously updated by the slower, absolute sensors (star cameras) and is robust against sensor failure. Optical encoders report the relative position of the inner frame to the outer frame. Motion sensing for the inner frame is provided by two sets of three orthogonally mounted, high bandwidth gyroscopes.

The absolute pointing sensors are two integrating star cameras [26] that are mounted above the receiver on the inner frame. Each star camera has an internal computer that calculates a real-time pointing solution at 1 Hz by comparing measured star separations with an on-board catalog of stars. The cameras are capable of dead reckoning. The two cameras run independently, providing failsafe redundancy. The camera system has been tested extensively on the ground and has flown five times on balloon payloads (four times on BLAST and once on the x-ray telescope InFOCμS). A comparison of simultaneous pointing solutions from both cameras gives a rms uncertainty of <2". To meet the absolute pointing requirements for TIM, we will reduce the field of view of the cameras by a factor of two and use new CCDs with enhanced quantum efficiency that roughly doubles the sensitivity in the far red (where TIM uses them). This will allow us to obtain high-accuracy, continuously updated pointing solutions for our observations.

### III. Flight Operations

TIM will perform its science from an Antarctica long duration balloon (LDB) flight. The richest field available from an Antarctic LDB flight is GOODS-S (03hr32m30s, -27d48m17s), the same field covered by BLAST in 2006 (Devlin et al. 2009). This field has deep coverage at optical through mm wavelengths. It is the target for the deep fields with HST, Chandra, Spitzer, Herschel, ALMA, and planned surveys with JWST and Euclid. There exist thousands of spectroscopic redshifts in the field, making it optimal for multi-wavelength counterpart identification and stacking analyses. For sun avoidance, we will need to choose two fields, roughly 12 hours in RA apart. For optimal survey strategy for a power spectrum analysis for the clustering term, this field should be wide and narrow to access the low k modes in Figure 4A. For pointing reconstruction, counterpart identification, line extraction, stacking analyses, and cross correlations we will want as much ancillary data as possible. We will place our second field in the SPT-Deep field (23hr30m00s -55d00m00s) which has Herschel/SPIRE, Spitzer/IRAC, XMM, radio, and optical data. Maps, catalogs, and raw data will be made publicly available to the community to increase the legacy value of the TIM data set.

We plan for a mission length of two weeks, for a total time aloft of 336 hours. Significantly longer flight times have been achieved; up to 28 days could be used given the cryostat hold time. We assume 2 hours in every 40 are taken up with recycling of the $^3$He fridges and other overhead operations, and ~10% of the time is given to calibration and pointing observations. This leaves 280 hours to be divided between two fields, for an approximate integration time per field of ~140 hours. Our current baseline plan is to map a small deep field (0.1 deg$^2$) in GOODS-S and a wider, shallower field (~1 deg$^2$) within the SPT Deep field.

The TIM scan strategy will be very similar to that of BLAST, with scans at fixed elevation while the telescope and spectrometer slits are moved back and forth in azimuth. The inner frame is then stepped to cover the entire field, by a larger amount for GOODS-S and smaller for SPT. The nominal offset between the star camera and spectrometer slits is determined on the ground. In flight, the offset is determined on bright targets. In both cases the approach is to map the source via small raster scans. The accuracy of the mechanical alignment and the blind pointing solution from the star cameras should be better than 1', easily allowing the target to appear in the small field map. The spectrometer data are summed spectrally to produce a continuum flux and reduced in real-time to locate the source. This offset will need to be periodically re-checked throughout the flight. These bright sources will also provide calibration of the PSF and focus checks for the secondary mirror.

The flight software for TIM will be designed so that it can operate autonomously after launch. The survey fields will be decided before the flight. The autonomous scheduling system (developed for BLAST) will use schedule files that consist of a sequential list of observations or actions as a function of the Local Sidereal Time. This system is robust against temporary system failures because the telescope only needs to know the current time and location to resume operation upon recovery. Using a local sidereal clock rather than a clock fixed in some time zone, it is possible to account for purely astronomical visibility constraints (such as the RA of the Sun and of the astronomical targets) using a static description. For every launch opportunity, six schedule files are generated, which account for 3 different cases of flight latitude and longitude, and two cases of measured instrument sensitivity. The gondola uses the GPS to decide which schedule file to use, appropriate for the declination of the target field. At the beginning of the flight, the sensitivities and beam size are estimated from scans across calibrators. Based on this information the ground station team can decide which of the two sets of schedule files the instrument should use, and switch between the two using a single command.

We will use several methods for primary calibration, most based on the successful approaches used by the direct-detection millimeter-wave spectrometer Z-Spec (e.g., [27], [28]). The TIM bands are sufficiently wide that a continuum calibrator can be used to calibrate both the absolute and relative response of the channels, and channels may also be co-added. For frequency calibration we will begin with a Fourier transform spectrometer (FTS) as a laboratory calibrator, as was done for Z-Spec. Observations of sources with rich, known spectra, such as evolved stars like NGC7027 [29], [30] may be used as spectral templates. A similar technique was used successfully for Z-Spec using IRC+10216. For TIM, we have the additional frequency calibration scale of line emission from the atmosphere itself (since these lines will be narrow and do not suffer the severe pressure broadening present in ground-based observations). The telescope modulation as well as telluric-line-dominated spectral channels may also be used to reduce any effects of time-varying atmospheric emission.

The data rate from the TIM detectors will be substantial, but not prohibitive. Even at the full sample rate of the readout, 488 Hz, the 3600 detectors sampled at 32 bits produces a data rate of 7 MB/s, or 25 GB per hour. Over the course of a 14-day flight, this results in a total data volume less than 10 TB, even including overheads for housekeeping data.

REFERENCES

[1] Madau, P. & Dickinson, M. 2014, ARAA, 52, 415
[2] Casey, C. M., Narayanan, D., & Cooray, A. 2014, Phys. Rep., 541, 45
[3] Hauser, M. G. & Dwek, E. 2001, ARAA, 39, 249
[4] Lagache, G., Puget, J.-L., & Dole, H. 2005, ARAA, 43, 727
[5] Lupu, R. E. et al. 2012, ApJ, 757, 135
[6] Barber, G. J. et al. 2008, Nuclear Instruments and Methods in Physics Research A, 593, 624
[7] Martin, R. N., Romeo, R. C., & Kingsley, J. S. 2006, in Proc. SPIE, Vol. 6273, Society of Photo-Optical Instrumentation Engineers (SPIE) Conference Series, 62730P
[8] Woody, D. et al. 2008, in Proc. SPIE, Vol. 7018, Advanced Optical and Mechanical Technologies in Telescopes and Instrumentation, 70180T
[9] Su, P., Parks, R. E., Wang, L., Angel, R. P., & Burge, J. H. 2010, in International Optical Design Conference and Optical Fabrication and Testing (Optical Society of America), OTuB3
[10] Pascale, E. et al. 2012, in Society of Photo-Optical Instrumentation Engineers (SPIE) Conference Series, Vol. 8444, Society of Photo-Optical Instrumentation Engineers (SPIE) Conference Series
[11] Pascale, E. et al. 2008, ApJ, 681, 400
[12] Ferkinhoff, C. et al. 2014, ApJ, 780, 142
[13] Parshley, S. C., Ferkinhoff, C., Nikola, T., Stacey, G. J., Ade, P. A., & Tucker, C. E. 2012, in Society of Photo-Optical Instrumentation Engineers (SPIE) Conference Series, Vol. 8452, Society of Photo- Optical Instrumentation Engineers (SPIE) Conference Series
[14] Dober, B. et al. 2016, Journal of Low Temperature Physics, 184, 173
[15] Galitzki, N. et al. 2016, in Proc. SPIE, Vol. 9914, Millimeter, Submillimeter, and Far-Infrared Detectors and Instrumentation for Astronomy VIII, 99140J
[16] McKenney, C. M., Leduc, H. G., Swenson, L. J., Day, P. K., Eom, B. H., & Zmuidzinas, J. 2012, in Society of Photo-Optical Instrumentation Engineers (SPIE) Conference Series, Vol. 8452, Society of Photo-Optical Instrumentation Engineers (SPIE) Conference Series
[17] Swenson, L. J. et al. 2012, in Society of Photo-Optical Instrumentation Engineers (SPIE) Conference Series, Vol. 8452, Society of Photo-Optical Instrumentation Engineers (SPIE) Conference Series
[18] Hailey-Dunsheath, S. et al. 2014, Journal of Low Temperature Physics, 1
[19] Hailey-Dunsheath, S. et al. 2016, Journal of Low Temperature Physics, 184, 180
[20] Shirokoff, E. et al. 2014, Journal of Low Temperature Physics, 1
[21] Wheeler, J. et al. 2016, in Proc. SPIE, Vol. 9914, Millimeter, Submillimeter, and Far-Infrared Detectors and Instrumentation for Astronomy VIII, 99143K
[22] Hubmayr, J. et al. 2013, Applied Superconductivity, IEEE Transactions on, 23, 2400304
[23] Gao, J. et al. 2008, Applied Physics Letters, 92, 212504
[24] Zmuidzinas, J. 2012, Annual Review of Condensed Matter Physics, 3, 169
[25] Hailey-Dunsheath, S. et al. 2018, Journal of Low Temperature Physics, submitted, arXiv/1803.02470
[26] Rex, M., Chapin, E., Devlin, M. J., Gundersen, J., Klein, J., Pascale, E., &Wiebe, D. 2006, in Presented at the Society of Photo-Optical Instrumentation Engineers (SPIE) Conference, Vol. 6269, Groundbased and Airborne Instrumentation for Astronomy. Edited by McLean, Ian S.; Iye, Masanori. Proceedings of the SPIE, Volume 6269, pp. 62693H (2006).
[27] Kamenetzky, J. et al. 2011, ApJ, 731, 83
[28] Naylor, B. J. et al. 2010, ApJ, 722, 668
[29] Groenewegen, M. A. T. et al. 2011, A&A, 526, A162
[30] Wesson, R. et al. 2011, in Astronomical Society of the Pacific Conference Series, Vol. 445, Why Galaxies Care about AGB Stars II: Shining Examples and Common Inhabitants, ed. F. Kerschbaum, T. Lebzelter, & R. F. Wing, 607